%% file: main.tex
\documentclass[conference]{IEEEtran}

\usepackage[T1]{fontenc}
\usepackage[utf8]{inputenc}

\usepackage{cite}
\usepackage{amsmath}
\usepackage{amssymb}
\usepackage{graphicx}
\usepackage{booktabs}
\usepackage{array}
\usepackage{multirow}
\usepackage{algorithm}
\usepackage[noend]{algpseudocode}
\usepackage{tikz}
\usetikzlibrary{arrows.meta,positioning,calc,fit}
\usepackage{xcolor}
\usepackage[hidelinks]{hyperref} 

\newcommand{\sys}{\mbox{ABCoord}}
\newcommand{\fit}{\mathit{f{}it}} 
\newcommand{\trial}{\mathit{trial}}
\newcommand{\SN}{\mathit{SN}}
\newcommand{\MCN}{\mathit{MCN}}

\begin{document}
\bstctlcite{IEEEexample:BSTcontrol}

\title{ABCoord: An Artificial Bee Colony (ABC)-Based Decentralized Distributed Two-Layer Decision-Making Architecture for the Coordination of AI Agents}

\author{\IEEEauthorblockN{D. Karaboga and B. A. Uluırmak}
\IEEEauthorblockA{\textit{Department of Computer Engineering, Faculty of Engineering}\\
\textit{Artificial Intelligence and Big Data Application and Research Center Laboratory}\\
\textit{Erciyes University}, Kayseri, Türkiye\\
karaboga@erciyes.edu.tr, uluirmak@erciyes.edu.tr}}

\maketitle

\input{sections/00_abstract}
\input{sections/01_introduction}
\input{sections/02_related_work}
\input{sections/03_background_abc}
\input{sections/04_architecture}
\input{sections/05_results_discussion}
\input{sections/09_conclusion}

\bibliographystyle{IEEEtran}
\bibliography{refs}

\end{document}

%% file: sections/00_abstract.tex
\begin{abstract} Rapid progress in large language models (LLMs) is turning AI agents into multi-agent systems that jointly
solve complex tasks, and the architecture that coordinates the agents toward a common goal is gaining fundamental
importance in these systems. Since centralized architectures can become bottlenecks as the number of agents grows, whereas
distributed architectures incur increasing communication costs, a decentralized coordination architecture that scales
with low communication cost is needed. Recently, the structure of the artificial bee colony (ABC) algorithm has been
defined for the first time as a two-layer decentralized information-processing and decision-making architecture.
\sys{} is built on this proposed architecture. The results show that \sys{}, used for the coordination of AI agents, performs at a level comparable to existing architectures on the BigCodeBench benchmark in the self-test
condition, in which the tests are also written by the language model. In the oracle condition, in which the solutions
are evaluated with the official tests, the number of solved problems increases considerably.
\end{abstract}


%% file: sections/01_introduction.tex
\section{Introduction}
\label{sec:intro}

Rapid progress in large language models is transforming AI agents from tools that work in isolation into multi-agent
systems that jointly solve complex tasks. As the number of agents and the diversity of tasks in these systems grow, a
fundamental problem arises: how autonomous agents coordinate task allocation, information exchange and their decisions
toward a common goal~\cite{wooldridge2009introduction}. A lack of coordination leads to duplicated work, conflicting
decisions, unnecessary communication overhead and loss of efficiency. This difficulty becomes more pronounced when
the environment and tasks change dynamically and the number of agents grows.

The coordination architectures proposed so far fall broadly into two groups. The first group comprises centralized and
hierarchical architectures in which a manager or orchestrator agent decomposes the tasks and distributes them to the
other agents~\cite{hong2024metagpt,huang2023agentcoder}. These architectures constitute a single point of failure and
can become bottlenecks as the number of agents grows~\cite{wooldridge2009introduction}. The second group comprises
distributed architectures in which the agents reach a joint decision through direct messaging, debate, voting or
auction-based protocols. In these approaches, however, the communication cost and the coordination complexity grow with
the number of agents. Therefore, there is a need for a decentralized and distributed decision-making coordination architecture
that requires no central control and scales with low communication cost.

Social organisms in nature can achieve a high degree of coordination without any central control, which makes swarm
intelligence-based algorithms a natural source of inspiration for this need. The most widely studied swarm intelligence
algorithms in the literature are particle swarm optimization (PSO)~\cite{eberhart1995new}, ant colony optimization
(ACO)~\cite{colorni1992distributed} and the artificial bee colony (ABC)~\cite{karaboga2005idea} algorithms. Until
recently, the ABC algorithm was mainly used to solve parametric optimization problems, and dozens of improved
versions of it were developed~\cite{akay2021survey}. From the perspective of how collective intelligence emerges, ABC differs considerably from PSO and ACO. Karaboga et
al.~\cite{karaboga2026stigmergy} showed that ants mainly use stigmergic information sharing when solving the foraging
problem, whereas honey bees combine direct information sharing in the hive with rapidly changing micro-stigmergic
information sharing on the flowers in the field. The joint operation of these two mechanisms makes it possible to define
the structure of ABC as a two-layer structure. Building on this observation, Karaboga et al.~\cite{karaboga2026stigmergy}
defined the foraging behavior of real honey bees, and hence the structure of the ABC algorithm, for the first time as a
two-layer, decentralized and distributed information-processing and decision-making architecture, and stated that this
architecture can be used for the self-coordination of artificial collectives.

In this study, this architecture (\sys{}) is used for the first time as an architecture for the coordination of AI agents. The contribution of the study is the adaptation of \sys{} to AI agents and the comparison of its performance
with existing coordination architectures in the literature on the BigCodeBench~\cite{zhuo2025bigcodebench} benchmark
problem set. The results show that \sys{} can be used successfully for the coordination of AI agents. These results
also indicate that the applicability of \sys{} to the coordination of other artificial collectives is worth
investigating in future studies.

%% file: sections/02_related_work.tex
\section{Related Work}
\label{sec:related}

Large language models (LLMs) are also used in program search methods. Among these methods,
FunSearch~\cite{romeraparedes2024mathematical} evolves programs sampled from a program database with an LLM, relying on
a given evaluator. The ABC algorithm, on the other hand, was previously applied to program trees through the ABC
programming~\cite{karaboga2012artificial} approach. More recently, it has also been used to search for ways of
circumventing the safety constraints of a language model over multiple turns (jailbreak)~\cite{liu2025bees}. However,
coordination among agents is not addressed in these studies.

Among the multi-agent systems developed for programming, AgentCoder~\cite{huang2023agentcoder} and
MetaGPT~\cite{hong2024metagpt} follow a fixed role script. In these systems, the script determines
which agent runs and when. By contrast, repeated sampling (Best-of-$N$)~\cite{brown2024large} scales inference-time computation
solely by increasing the number of sampled solutions. Generated candidate solutions can also be checked against an
LLM-written reference solution~\cite{zhang2023algo}.

%% file: sections/03_background_abc.tex
\section{Artificial Bee Colony Algorithm}
\label{sec:background}

The general scheme of the ABC algorithm~\cite{karaboga2005idea}, as given by Karaboga~\cite{karaboga2010scholarpedia}, is
shown in Algorithm~\ref{alg:abc}. The colony works on $\SN$ food sources $x_i$. Each source represents a candidate
solution and carries a fitness value $\fit_i$ and a trial counter $\trial_i$. One half of the colony consists of
employed bees (one employed bee per source) and the other half of onlooker bees. An employed bee produces a new solution
in the neighborhood of its own source as \begin{equation} v_{ij} = x_{ij} + \phi_{ij}\,(x_{ij} - x_{kj})
\label{eq:neighbour} \end{equation} where $k \neq i$ denotes another randomly selected source, $j$ a randomly selected
dimension and $\phi_{ij}$ a uniformly distributed number in $[-1,1]$. In greedy selection, if $\fit(v_i) > \fit(x_i)$,
$v_i$ is accepted and $\trial_i$ is reset. Otherwise, $\trial_i$ is incremented by one. An onlooker bee selects a source
with probability \begin{equation} p_i = \fit_i \Big/ \textstyle\sum_{m=1}^{\SN} \fit_m \label{eq:roulette}
\end{equation} that is, by roulette selection, and applies the employed bee step to this source. A source whose trial
counter exceeds the limit $L$ is abandoned. The scout bee replaces this source with a new solution generated randomly
from the search space.

\begin{algorithm}[H]
\caption{Standard ABC algorithm.}
\label{alg:abc}
\footnotesize
\begin{algorithmic}[1]
\State Initialization phase
\Repeat
  \State Employed bee phase
  \State Onlooker bee phase
  \State Scout bee phase
  \State Memorize the best solution achieved so far
\Until{the cycle count reaches $\MCN$ or the maximum CPU time is reached}
\end{algorithmic}
\end{algorithm}

%% file: sections/04_architecture.tex
\section{\sys{} Architecture}
\label{sec:architecture}

The \sys{} architecture uses the phases and decision rules of the standard ABC algorithm without modification. Only the
counterparts of the food source, fitness, neighbor solution and bee concepts change. Each food source represents a
candidate solution consisting of Python code that carries out the task. The fitness value is the fraction of tests the
solution passes (Section~\ref{sec:fitness}). A neighbor solution corresponds to a small edit made to this code with the
LLM, and each bee to a call to the same LLM with its own role prompt.

In their recent work, Karaboga et al.~\cite{karaboga2026stigmergy} defined the structure of the artificial bee colony for
the first time as a two-layer decentralized information-processing and decision-making architecture. \sys{} is
built on this proposed architecture (Fig.~\ref{fig:katman}). In the hive, which is the strategic layer, the fitness
values of the sources are shared, and the onlooker bees select the source they will work on by roulette selection based
solely on these values. In the field, which is the operational layer, each source holds the candidate solution, the
tests this solution fails and the trial counter. The bees edit by reading this information at the source and
influence one another indirectly by changing it.

\begin{figure}[H]
\centering
\input{figures/fig_katman}
\caption{Two-layer structure of foraging in ABC (Karaboga et al.~\cite{karaboga2026stigmergy}).}
\label{fig:katman}
\end{figure}

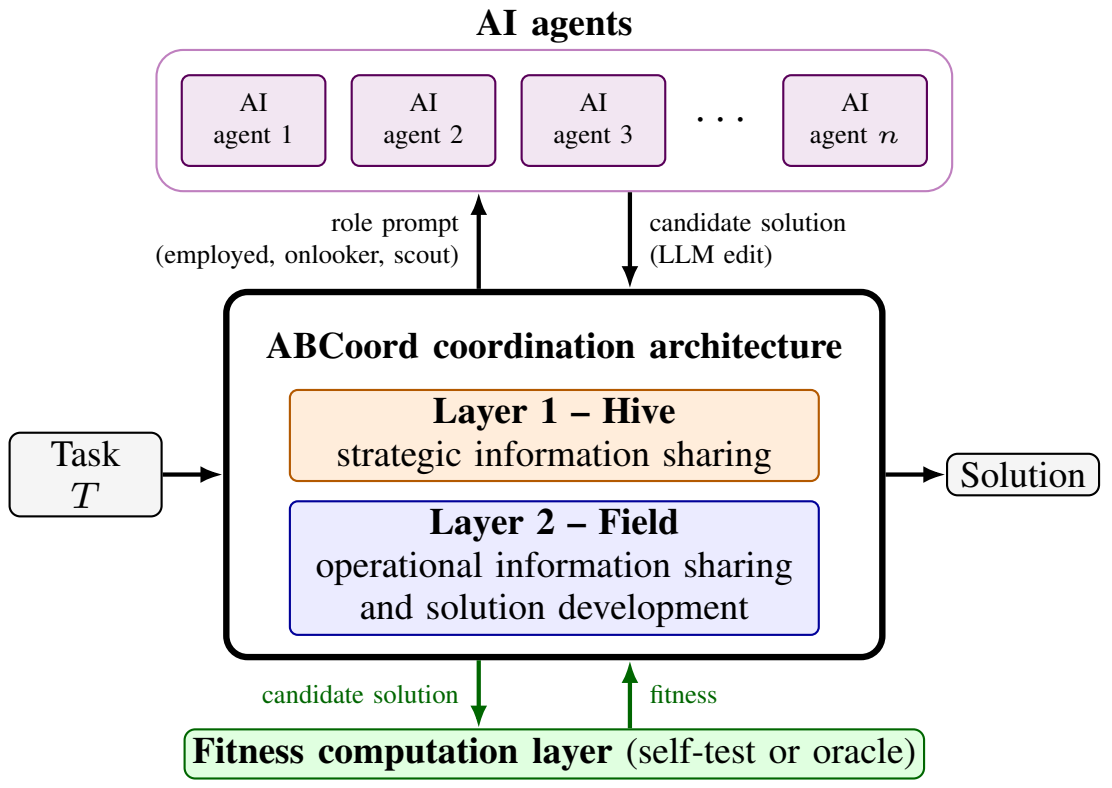
\begin{figure}[H] \centering \input{figures/fig_blok} \caption{Block diagram of \sys{}.} \label{fig:blok} \end{figure}

As seen in Fig.~\ref{fig:blok}, there is neither a manager agent nor messaging among the AI agents. The two-layer
\sys{} architecture decides with which role and for which candidate solution each LLM call is made. All state
information is kept in the colony, and each call receives only the information required by its role in ABC. Every
decision of the colony is also recorded. The cost of each call is deducted from a single token budget $B$ for the task.

In standard ABC, a solution is a vector of numbers, so a new solution is generated by selecting random numbers. In
\sys{}, a solution is Python code, so the initial solutions are generated by having the LLM first write a short plan and
then the code. To prevent the solutions from being identical, the model is shown the approaches followed before and is
asked to follow a different one. The scout bee applies the same step when opening a new source.

In Eq.~\eqref{eq:neighbour}, the counterpart of $x_{ij}$ in \sys{} is the current solution $x_i$. The counterpart of the
difference term is a \emph{hint} taken from another randomly selected source $x_k$. This hint consists of the lines in
which the code of the two solutions differs or an observed behavior of $x_k$, and it is given to the model together with
the tests that $x_i$ fails. The counterpart of changing a single dimension $j$ is keeping the edit small. Every accepted
edit is added to the archive.

\begin{algorithm}[H]
\caption{\sys{} algorithm.}
\label{alg:abcoord}
\footnotesize
\begin{algorithmic}[1]
\State Different initial solutions (food sources) are written for the task, and the fraction of tests each solution passes (fitness) is computed
\Repeat
  \State In the employed bee phase, each bee makes a small edit to its own solution with the LLM, and the edited solution replaces the old one if it passes more tests
  \State In the onlooker bee phase, solutions with higher fitness are selected with higher probability, and the same edit step is applied
  \State The fitness values of the solutions are recomputed
  \State In the scout bee phase, a solution that has not been improved for a long time is abandoned and a new solution is written in its place
\Until{the maximum cycle count is reached, the colony can no longer progress or the token budget is exhausted}
\State The best solution found is delivered
\end{algorithmic}
\end{algorithm}

As seen in Algorithm~\ref{alg:abcoord}, \sys{} runs the employed bee, onlooker bee, fitness update and scout bee phases in
sequence. Since the onlooker bees run one after another, each of them continues from the edits already accepted in the
same cycle. Because fitness is a pass rate, it can be 0 and saturates at 1. In addition, tests are added during the
search, and more than one bee can work on a source in the same cycle. The rules of ABC have been adapted to these
situations. A small term $\varepsilon=0.05$ added to roulette selection keeps every source selectable. A saturated
source ($\fit_i=1$) is left untouched, and the trial counter is incremented at most once per cycle. When all sources
are saturated and new tests can no longer distinguish them, the search is stopped early. The best solution is selected
from the archive according to the final tests.

%% file: figures/fig_katman.tex
\begin{tikzpicture}[font=\scriptsize, >={Latex[length=1.5mm]},
  src/.style={draw=blue!60!black, fill=blue!8, rounded corners=1.5pt, minimum width=13mm, minimum height=7mm},
  role/.style={anchor=north west, align=left, text width=84mm, inner sep=0.6pt, font=\footnotesize}]
\node[draw=orange!70!black, fill=orange!14, rounded corners=2pt, minimum width=20mm, minimum height=8mm, align=center,
  font=\footnotesize] (hive) at (-6mm,0) {Dance};
\node[anchor=south, font=\scriptsize\bfseries, inner sep=1pt] at (hive.north) {Hive};
\node[anchor=west, align=left, text=orange!50!black] at ($(hive.east)+(2.5mm,0)$)
  {\textbf{Layer 1} -- strategic,\\ direct information sharing};
\node[src] (s1) at (-30mm,-21mm) {};
\node[src] (s2) at (-21mm,-18mm) {};
\node[src] (s3) at (-16mm,-24mm) {};
\node[font=\footnotesize] (dd) at (-2mm,-21mm) {$\cdots$};
\node[src] (sn) at (13mm,-21mm) {};
\foreach \s/\t in {s1/1, s2/2, s3/3, sn/N} {\node[font=\footnotesize\bfseries, text=blue!50!black] at (\s) {\t};}
\draw[densely dashed, blue!40, rounded corners=4pt] ($(s1.west)+(-1.5mm,7.5mm)$) rectangle ($(sn.east)+(1.5mm,-7mm)$);
\node[anchor=west, align=left, font=\scriptsize\bfseries, inner sep=1pt] at ($(sn.east)+(3mm,0)$) {Floral\\ regions};
\node[anchor=north, align=center, text=blue!50!black] at (-8mm,-29.5mm)
  {\textbf{Layer 2} -- operational area, stigmergic information sharing\\ and solution development field};
\draw[<->, thick] ($(hive.south)+(-8mm,0)$) -- (s1.north);
\draw[<->, thick] ($(hive.south)+(-3mm,0)$) -- (s2.north);
\draw[<->, thick] ($(hive.south)+(7mm,0)$) -- (sn.north);
\node[anchor=north east, align=left, inner sep=0pt] at (-25.5mm,-0.5mm) {\textbf{Bees}\\ employed bee\\ onlooker bee\\ scout bee};
\node[role] (r1) at (-41mm,-37.5mm) {An \textbf{employed bee} makes a small LLM edit on its own source (solution); if the edit has
  higher fitness, it accepts the edit and shares the fitness of the source in the hive.};
\node[role, below=0.6mm of r1.south west, anchor=north west] (r2) {An \textbf{onlooker bee} selects a source by the roulette
  wheel method according to the fitness values shared in the hive and makes an edit as in the employed bee phase.};
\node[role, below=0.6mm of r2.south west, anchor=north west] (r3) {An employed bee whose source's trial counter exceeds the limit
  abandons the source, becomes a \textbf{scout bee} and discovers an entirely new solution.};
\node[role, below=0.6mm of r3.south west, anchor=north west] (r4) {If the bees cannot improve a source, they increment its trial
  counter by one, that is, they leave a \textbf{trace} at the source.};
\end{tikzpicture}

%% file: figures/fig_blok.tex
\begin{tikzpicture}[font=\scriptsize, >={Latex[length=1.5mm]},
  blk/.style={draw, rounded corners=2pt, align=center, inner sep=1.5pt, fill=gray!8, text width=9mm},
  llm/.style={draw=violet!70!black, fill=violet!10, rounded corners=1.5pt, minimum width=9.5mm, minimum height=6mm,
              align=center, inner sep=0.8pt, font=\tiny},
  lay/.style={draw, rounded corners=1.5pt, text width=34mm, align=center, inner sep=1.4pt}]
\node[llm] (l1) at (0,0) {AI\\agent 1};
\node[llm, right=1.6mm of l1] (l2) {AI\\agent 2};
\node[llm, right=1.6mm of l2] (l3) {AI\\agent 3};
\node[right=0.6mm of l3, font=\footnotesize] (dots) {$\cdots$};
\node[llm, right=0.6mm of dots] (ln) {AI\\agent $n$};
\node[draw=violet!50, rounded corners=5pt, fit=(l1)(ln), inner sep=1.6mm,
  label={[font=\scriptsize\bfseries, anchor=south, inner sep=1pt]north:AI agents}] (agents) {};
\node[font=\scriptsize\bfseries, below=8mm of agents.south] (title) {\sys{} coordination architecture};
\node[lay, fill=orange!14, draw=orange!70!black, below=0.8mm of title] (hive)
  {\textbf{Layer 1 -- Hive}\\ strategic information sharing};
\node[lay, fill=blue!8, draw=blue!60!black, below=1.2mm of hive] (field)
  {\textbf{Layer 2 -- Field}\\ operational information sharing\\ and solution development};
\node[draw, very thick, rounded corners=4pt, fit=(title)(hive)(field), inner sep=1.4mm] (abc) {};
\node[blk, left=4mm of abc] (task) {Task\\ $T$};
\node[blk, right=4mm of abc] (out) {Solution};
\draw[->, thick] (task) -- (abc);
\draw[->, thick] (abc) -- (out);
\node[draw, rounded corners=2pt, fill=green!12, draw=green!40!black, below=4.5mm of abc, align=center, inner sep=1.5pt]
  (fitl) {\textbf{Fitness computation layer} (self-test or oracle)};
\draw[->, thick] ($(abc.north)+(-5mm,0)$) -- node[left, font=\tiny, align=right] {role prompt\\(employed, onlooker, scout)} ($(agents.south)+(-5mm,0)$);
\draw[->, thick] ($(agents.south)+(5mm,0)$) -- node[right, font=\tiny, align=left] {candidate solution\\(LLM edit)} ($(abc.north)+(5mm,0)$);
\draw[->, thick, green!40!black] ($(abc.south)+(-5mm,0)$) -- node[left, font=\tiny] {candidate solution} ($(fitl.north)+(-5mm,0)$);
\draw[->, thick, green!40!black] ($(fitl.north)+(5mm,0)$) -- node[right, font=\tiny] {fitness} ($(abc.south)+(5mm,0)$);
\end{tikzpicture}

%% file: sections/05_results_discussion.tex
\section{Results and Discussion}
\label{sec:sonuclar}

\subsection{Fitness Computation Layer}
\label{sec:fitness}

In \sys{}, the fitness value required by ABC is computed through a single interface. Thus, the search loop remains
unchanged whether the self-test or the oracle fitness signal is used.

In the self-test condition, the tests are written by the colony. When the existing tests are insufficient, one of the
onlooker bees takes on the role of \emph{test bee} during the fitness update. Without seeing the candidate code, this bee
determines the test inputs and the expected outputs by looking at the task description and the execution summary of the
inputs it has written before. The fitness of a solution $x$ is the weighted pass rate over these tests and is computed as
\begin{equation} \fit(x) = \frac{\sum_t w_t\, g_t(x)}{\sum_t w_t} \label{eq:fitness} \end{equation}
where $w_t$ denotes the weight of test $t$, determined according to its reliability, and $g_t(x)$ is 1 if solution $x$
passes test $t$ and 0 otherwise.

In the oracle fitness condition, $\fit$ is the fraction of the official benchmark tests that the solution passes. The
acceptance, roulette selection, stopping and delivery decisions are based on this fraction. Failures on the tests written
by the test bee continue to guide the edits. However, no official test or score is shown to the model. Since the
decisions rely on the official tests, the results of this condition are not the performance of a ready-to-use
system but a diagnostic of how far the loop can progress when the fitness signal is correct.

\subsection{Experimental Setup}
\label{sec:setup}

The experiments use 232 problems from the Instruct split of the BigCodeBench~\cite{zhuo2025bigcodebench} benchmark.
These problems require the combined use of common Python libraries and are evenly distributed over three difficulty
levels defined in this study. For each problem, the single delivered solution is tested with the official unit tests in
a container without network access, and the number of solved problems is used as the performance measure. All agents use
the Qwen3-Coder~30B~\cite{qwen2025qwen3coder} model served locally via Ollama. In the self-test condition, \sys{} was run
with $\SN=3$ (six bees), $L=5$ and $\MCN=20$. In the oracle condition, $\SN=5$ (ten bees) and $L=3$ were used. In this
condition, the scout bees generate new solutions without being shown the previous approaches, the trial counter is
incremented after every failed attempt, and the colony stops when it finds a solution that passes all official tests. The
Best-of-$N$~\cite{brown2024large}, AgentCoder~\cite{huang2023agentcoder} and
FunSearch~\cite{romeraparedes2024mathematical} methods were re-implemented in this study by adapting them to the same
environment, and they use the same model, container and evaluation procedure. In the oracle condition, all
architectures were given an equal budget of 160k tokens per problem.

\subsection{Results}
\label{sec:results}

Table~\ref{tab:main} compares the four architectures in the two conditions and gives the average token consumption per
problem for each condition. In both conditions, \sys{} yields results at a level comparable to the existing
architectures. In the oracle fitness condition, the number of problems solved by \sys{} increases from 122 to 160.

\begin{table}[H] \caption{Number of solved problems and average tokens per problem.}
\label{tab:main} \centering \footnotesize \setlength{\tabcolsep}{4pt}
\begin{tabular}{@{}lcccc@{}} \toprule
 & \multicolumn{2}{c}{Self-test fitness} & \multicolumn{2}{c}{Oracle fitness} \\ \cmidrule(lr){2-3} \cmidrule(l){4-5}
Architecture & Solved & Tokens & Solved & Tokens \\ \midrule
\sys{} (ABC) & 122 & 31.4k & 160 & 55.4k \\
Best-of-$N$ & 120 & 27.3k & 151 & 59.7k \\
AgentCoder & 112 & 17.6k & 155 & 60.3k \\
FunSearch & 117 & 29.4k & 146 & 63.6k \\
\bottomrule \end{tabular} \end{table}

\begin{figure}[H]
\centering
\includegraphics[width=\columnwidth]{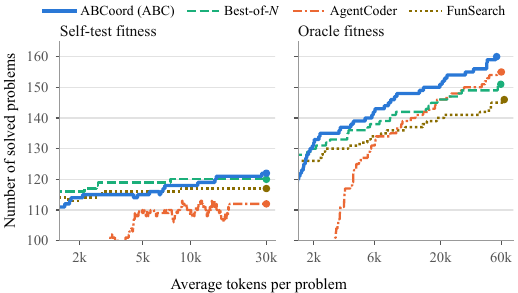}
\caption{Number of solved problems versus average tokens per problem in the two conditions.}
\label{fig:gelisim}
\end{figure}

Both the selection among the initial solutions and the search contribute to this increase. The oracle signal selects the
correct one among the initial solutions. The cycle formed by the employed, onlooker and scout bees, on the other hand, produces
correct solutions for problems where none of the initial solutions is correct. Fig.~\ref{fig:gelisim} shows the number of
problems solved by the four architectures against the average tokens per problem in both conditions.
In the self-test condition, the curves level off after about 10k tokens, whereas in the oracle condition they keep
rising as token consumption increases.

\subsection{Discussion}
\label{sec:discussion}

In the self-test condition, the colony produces a correct solution for some problems and adds it to its archive, but it
cannot deliver this solution because it cannot distinguish it with its own tests. This indicates that the factor that
primarily limits performance in the ready-to-use system is the fitness signal rather than the search loop. A possible
reason for the inadequacy of the signal in this condition is that the tests are written by the same model that writes
the solutions. Since fitness is computed through a single interface, stronger signals, such as tests written by a
different model, can be used in the fitness computation layer without changing the colony. Nevertheless, it should be
noted that all results were obtained on a single benchmark with a single open-weight model and that the
architecture was developed on this problem set. The number of bees in the colony was kept quite small in this first
study. Performance analyses of the proposed architecture under different colony settings are left for future work.

%% file: sections/09_conclusion.tex
\section{Conclusion}
\label{sec:conclusion}

In this study, \sys{}, a two-layer, decentralized and distributed decision-making architecture, was used for the first
time for the coordination of AI agents. The experiments show that ABC can be applied to AI agents while its
cycle is preserved. In this adaptation, ABC is observed to operate as a coordination layer that relies on explicit rules and
records every decision. Solving 160 of the 232 problems in the oracle fitness condition shows the level the colony can
reach when the fitness signal is correct. Future work will use stronger fitness signals in the fitness computation layer.